\documentclass[aps,prl,twocolumn,showpacs,floatfix,twoside,superscriptaddress]{revtex4-2}
\usepackage{amssymb}
\usepackage{graphicx}
\usepackage{amsmath}
\usepackage{amsfonts}
\usepackage{bbold}
\usepackage{esint}
\usepackage{color}
\usepackage[dvipsnames]{xcolor}
\usepackage{float}
\usepackage{verbatim}
\usepackage[utf8]{inputenc}
\usepackage{verbatim}
\usepackage[colorlinks=true,citecolor=blue,linkcolor=blue,urlcolor=blue]{hyperref}
\usepackage[justification=centerlast,font={small}]{caption}

\DeclareMathOperator{\re}{Re}

\newcommand{\BN}{\mathcal{N}}

\begin{document}

 \title{Anomalous Knudsen effect signaling long-lived modes in 2D electron gases
 }
\author{Grigory A. Starkov}
\author{Bj\"orn Trauzettel}
\affiliation{Institute for Theoretical Physics and Astrophysics,
University of W\"urzburg, D-97074 W\"urzburg, Germany}
\affiliation{W\"urzburg-Dresden Cluster of Excellence ctd.qmat, Germany}
\email{grigorii.starkov@uni-wuerzburg.de}

\date{\today}

\begin{abstract}
 Proper analysis of electron collisions in two spatial dimensions leads to the conclusion, that the odd harmonics of the electron distribution function decay much slower than the even ones. At the same time, the number of long-lived odd harmonics quickly shrinks with increasing temperature. Focusing on a channel geometry with boundary scattering, we show that such behavior of the odd decay rates leads to a characteristic behaviour of the conductance that we dub anomalous Knudsen effect: it initially grows with temperature but then starts to decrease, forming a peak.
 Further increase of the temperature forces the conductance to grow again due to the Gurzhi effect, associated with the crossover from ballistic to hydrodynamic transport.
 The simultaneous observation of the Gurzhi dip preceded by the anomalous Knudsen peak
 constitutes a particular signature of the long-lived modes in 2D electron transport at low temperatures.
\end{abstract}

\maketitle

\textit{Introduction\/.}--- Spearheaded by the emergence of ultra-clean two-dimensional materials such as Ga(Al)As heterostructures, graphene, or transition-metal dichalcogenides (TMDCs), recent years have seen substantial advances, both experimental~\cite{kukushkin_2004, weber_2005, bykov_2007, khodas_2010, dai_2010, bockhorn_2011, lee_2011, hatke_2012, mani_2013, yigen_2014, shi_2014, moll_2016, gooth_2018, sulpizio_2019, aharon-steinberg_2022} and theoretical~\cite{spivak_2006, muller_2008, mendoza_2013, narozhny_2015, lucas_2017, narozhny_2017, link_2018, alekseev_2018, alekseev_2019, holder_2019, raichev_2020, afanasiev_2021, zohrabyan_2023, heilmann_2024}, in the field of two-dimensional electron transport. The flow of electrons is determined by the interplay of electron-electron collisions with other types of scattering in the system, and an interesting part of the recent efforts~\cite{ledwith_2019, ledwith_2019b, ledwith_2019a,kryhin_2023, hofmann_2023, kryhin_2023a,kryhin_2025} is focused on the special features of electron-electron collisions that are characteristic of the two-dimensional geometry.

Electron-electron relaxation in 2D is dominated by head-on collisions. However, as first pointed out by Gurzhi and coauthors~\cite{gurzhi_1995}, it only affects the even part of the electron distribution function: The contribution of the head-on collisions to the relaxation of the odd part cancels out, and the subdominant scattering channel needs to be taken into account. As such, the odd harmonics become long-lived compared with the even ones.
This results in the appearance of the novel ``tomographic'' regime of the electron flow in two dimensions~\cite{ledwith_2019, ledwith_2019a}, characterized by particular flow profiles~\cite{ledwith_2019a} and scale-dependent viscosities~\cite{ledwith_2019a, kryhin_2023a, kryhin_2025}.

Although the spatial profile of the electron flow can already be observed experimentally~\cite{sulpizio_2019},
it is difficult to associate it with a given transport regime~\cite{heilmann_2024}.
It has also been argued, that the tomographic flow can explain the linear-in-temperature dependence of kinematic viscosity in highly-doped mono- and bi-layer graphene~\cite{zeng_2024}. In Ref.~\cite{moiseenko_2024}, it has been suggested that the lifetimes of the angular harmonics can be directly obtained from the shapes of cyclotron resonances. Although their measurements report longer lifetime of the third-order harmonic as of the second-order one, the difference is not large and results remain inconclusive.

Overall, looking for some simple signatures of the tomographic flow regime remains an open problem.
In our work, we predict a unique transport signature stemming from the long-lived modes in 2D.

The angular diffusion of the electrons on the Fermi surface can be described by the kernel $I_{ee}(\varphi,\varphi^\prime)$, where $\varphi$ and $\varphi^\prime$ parametrize two positions on it. If we assume the shape of the Fermi surface to be almost circular, we can expand the kernel in circular harmonics
\begin{equation}
 -I_{ee}(\varphi,\varphi^\prime) = \frac{1}{2\pi\tau_{ee}^{(0)}} + \sum_{n=1}^{+\infty} \frac{\cos{\left[n\left(\varphi-\varphi^\prime\right)\right]}}{\pi\tau_{ee}^{(n)}},\label{cint-expansion}
\end{equation}
introducing relaxation times $\tau_{ee}^{(n)}$ for different harmonics of the electron distribution function. Previous works~\cite{giuliani_1982, zheng_1996, ledwith_2019, ledwith_2019b, ledwith_2019a,kryhin_2023, hofmann_2023, kryhin_2023a,kryhin_2025} suggest that (for $k\in{0,1,2,\dotsc}$)
\begin{equation}
 \frac{1}{\tau_{ee}^{(2k)}}\equiv \frac{1-\delta_{k,0}}{\tau_{ee}^{(\mathrm{even})}} \cong (1-\delta_{k,0})\pi^2(T/E_F)^2 E_F/(2h),\label{tau-even}
\end{equation}
\begin{equation}
 \frac{1}{\tau_{ee}^{(2k+1)}}= \min\left\{\frac{(1-\delta_{k,0})}{\tau_{ee}^{(\mathrm{odd})}} (2k+1)^4, \frac{1}{\tau_{ee}^{(\mathrm{even})}}\right\},\label{tau-odd}
\end{equation}
while the ratio between even and odd relaxation times scales with the temperature as
\begin{equation}
 {\tau_{ee}^{(\mathrm{even})}}/{\tau_{ee}^{(\mathrm{odd})}} = (T/E_F)^2.\label{tau-ratio}
\end{equation}
Hence, the scattering rates $1/\tau^{(2k+1)}_{ee}$ for the odd harmonics grow with the order of the harmonic and saturate at the level of the even scattering rate for
\begin{equation}
 k^*\sim \left({\tau_{ee}^{(\mathrm{odd})}}/{\tau_{ee}^{(\mathrm{even})}}\right)^{1/4} = \sqrt{{E_F}/{T}}
\end{equation}
As the temperature grows, the number of the long-lived odd harmonics shrinks, which should have observable consequences.

In this paper, we study the temperature dependence of the electron flow induced by the applied homogeneous electric field both analytically and numerically.
We consider a narrow channel geometry with realistic boundary conditions in the high-mobility regime, where the impurity scattering length is larger than the width of the channel.
On the analytical side, we show up to first order in the electron collisions that the long-lived harmonics lead to an increase of the conductance with temperature, which agrees with Ref.~\cite{kryhin_2025}. Contrary to that, however, here we argue that the dependence on the temperature is manifestly non-linear: it is quadratic at first, but as the number of the long-lived harmonics dwindles with increasing temperature, the conductance quickly starts to decrease. We dub such unusual behavior anomalous Knudsen effect. When the temperature is increased even further, the system transitions eventually into the hydrodynamic regime, where the conductivity again grows due to the Gurzhi effect~\cite{gurzhi_1963, gurzhi_1968}.
This results in a characteristic picture, where the Gurzhi ``dip'' is preceded by an anomalous Knudsen ``peak''.
We argue that such behaviour can serve as the unique transport signature of the long-lived odd harmonics at low temperatures~\footnote{Let us mention the importance of the observation of both peak and dip together. At larger concentration of impurities, the conductance would also first increase even in the absence of long-lived modes. At higher temperatures it would decay due to the growing influence of the scattering on phonons, which would result in a peak of completely different nature.}.

\textit{Linearized Boltzmann Equation\/.}--- We consider an infinitly long channel of width $W$ confined to two spatial dimensions.
In the presence of an external homogeneous electric field $\vec E$ applied along the strip, the stationary flow of the electrons can be described by the linearized Boltzmann equation (assuming that the field is not very large):
\begin{equation}
 \vec v\cdot\vec\nabla_x \delta f + e\vec E\cdot \vec\nabla_p f_0 = -\frac{\delta f}{\tau_i} + I_{ee}\left[\delta f\right],\label{lbe-raw}
\end{equation}
where $\delta f$ is the deviation of the distribution function from its equilibrium value $f_0$. We have explicitly split the linearized collision operator into two parts: the first term on the right-hand side is attributed to the momentum-nonconserving scattering on impurities while the second term corresponds to the momentum-conserving electron-electron collisions. For the stationary flow, $\delta f\equiv\delta f_{\varepsilon\varphi}(x)$ depends only on the transverse coordinate $x$ across the strip. Here, $\varepsilon$ is the energy with respect to the chemical potential.

In addition to that, we take into account the scattering at the boundaries, characterizing it by an angular-dependent reflectivity parameter $r_{\varepsilon\varphi}$~\cite{soffer_1967, falkovskii_1970, okulov_1975, raichev_2020}:
\begin{multline}
\delta f_{\varepsilon\varphi}(0) = r_{\varepsilon\varphi}\delta f_{\varepsilon2\pi-\varphi}(0) \\ +\cfrac{1-r_{\varepsilon\varphi}}{\BN}\int_0^\pi d\varphi^\prime \sin{\varphi^\prime}(1-r_{\varepsilon\varphi^\prime})\delta f_{\varepsilon2\pi-\varphi^\prime}(0),\label{left-boundary}
\end{multline}
\begin{multline}
\delta f_{\varepsilon2\pi-\varphi}(W) = r_{\varepsilon\varphi}\delta f_{\varepsilon\varphi}(W) \\ +\cfrac{1-r_{\varepsilon\varphi}}{\BN}\int_0^\pi d\varphi^\prime \sin{\varphi^\prime}(1-r_{\varepsilon\varphi^\prime})\delta f_{\varepsilon\varphi^\prime}(W),\label{right-boundary}
\end{multline}
\begin{equation}
 \BN = \int_0^\pi d\varphi \sin{\varphi} (1-r_{\varepsilon\varphi}),
\end{equation}
where $\varphi\in[0,\pi]$.
Physically, Eqs.~\eqref{left-boundary} and~\eqref{right-boundary} imply that the electron either specularly reflects at the boundary with probability $r_{\varepsilon\varphi}$ or diffusively scatteres with probability $(1-r_{\varepsilon\varphi})$. Note that this form of the boundary conditions ensures that the current through the boundary vanishes.

Making the substitution $\delta f = -\frac{\partial f_0}{\partial \varepsilon} \eta$, we can use then the strongly peaked character of $\left(-\frac{\partial f_0}{\partial \varepsilon}\right)$ in the limit $T\ll E_F$ to integrate out $\varepsilon$ effectively pinning the energy to the Fermi surface $\varepsilon=0$. Under the assumption of an isotropic electron dispersion, the resulting equation for $\eta\equiv\eta_\varphi(x)$ takes the form
\begin{multline}
 \left[v_F \sin{\varphi}\frac{\partial}{\partial x} + \frac{1}{\tau_i}\right] \eta_\varphi(x) - eEv_F\cos{\varphi} \\= \int_0^{2\pi} d\varphi^\prime I_{ee}(\varphi-\varphi^\prime) \eta_{\varphi^\prime}(x).\label{lbe-projected}
\end{multline}
The kernel $\left(-I_{ee}(\varphi-\varphi^\prime)\right)$ describes the scattering of the electrons on the Fermi surface stemming from electron-electron collisions. Due to the assumption of isotropy it depends only on the difference of the ``out'' ($\varphi$) and ``in'' ($\varphi^\prime$) angles.
The expansion of the kernel in circular harmonics is specified in Eq.~\eqref{cint-expansion}.

Finally, we introduce the dimensionless coordinate $y=x/W-1/2$ ($y\in[-1/2,1/2]$) and make the substitution $\eta = eEW\xi$ to obtain the dimensionless representation of Eq.~\eqref{lbe-projected}:
\begin{equation}
 \left[\sin{\varphi}\frac{\partial}{\partial y} + \gamma_i\right]\xi_\varphi(y) = \cos{\varphi} + \int_0^{2\pi}d\varphi^\prime \tilde I_{ee}(\varphi-\varphi^\prime) \xi_{\varphi^\prime}(y),\label{final-lbe}
\end{equation}
\begin{equation}
 -\tilde I_{ee}(\varphi-\varphi^\prime) = \frac{\gamma_{ee}^{(0)}}{2\pi} + \sum_{n=1}^{+\infty}\frac{\gamma_{ee}^{(n)}}{\pi}\cos{\left[n(\varphi-\varphi^\prime)\right]},
\end{equation}
where
\begin{equation}
 \gamma_i = \frac{W}{v_F\tau_i}
 ,\qquad \gamma_{ee}^{(n)} = \frac{W}{v_F\tau_{ee}^{(n)}}
 \label{dless-fseries}
\end{equation}
are the dimensionless relaxation rates for scattering off impurities and for electron-electron collisions respectively. Since electron-electron scattering conserves the total momentum and the number of particles, $\gamma_{ee}^{(1)}=\gamma_{ee}^{(0)}=0$. Analogously, we can define the dimensionless even and odd relaxation rates
\begin{equation}
\gamma_{ee}^{(\mathrm{even})} = \frac{W}{v_F\tau_{ee}^{(\mathrm{even})}},\qquad \gamma_{ee}^{(\mathrm{odd})} = \frac{W}{v_F\tau_{ee}^{(\mathrm{odd})}}
\end{equation}
Focusing on the deviations
\begin{multline}
 \Delta \gamma_k = \gamma_{ee}^{(\mathrm{even})} - \gamma_{ee}^{(2k+1)} \\= \gamma_{ee}^{(\mathrm{even})}\left[1 - \min\{(2k+1)^4(T/E_F)^2,1\}\right]
\end{multline}
of the odd scattering rates from the even one, we can reformulate the kernel of the electron-electron collision operator in the form
\begin{multline}
 -\tilde I_{ee}(\varphi - \varphi^\prime)  = \gamma_{ee}^{(\mathrm{even})}\left[\delta_c(\varphi-\varphi^\prime)-1/(2\pi)\right] \\- \sum_{k=0}^{k^*} \frac{\Delta\gamma_k}{\pi} \cos{\left[(2k+1)(\varphi-\varphi^\prime)\right]},\label{deviation-form}
\end{multline}
where $\delta_c(\varphi-\varphi^\prime) = \sum_{n=-\infty}^{+\infty} \delta(\varphi-\varphi^\prime+2\pi n)$ is the Dirac delta-function on a circle.
Note that $\Delta \gamma_0=\gamma_{ee}^{(\mathrm{even})}$ due to momentum conservation.

\textit{Integral equation\/.}---
In order to compute the perturbative corrections in electron-electron collision operator in an iterative manner, it is convenient to transform the integro-differential equation~\eqref{final-lbe} into a purely integral equation similar to Refs.~\cite{dejong_1995, raichev_2020}.

Let us denote the right-hand side of Eq.~\eqref{final-lbe} by $\chi(y,\varphi)$.
Then, we can rewrite Eq.~\eqref{final-lbe} as
\begin{equation}
 \left[\frac{\partial}{\partial y} + \frac{\gamma_i}{\sin{\varphi}}\right]\xi(y,\varphi) = \frac{\chi(y,\varphi)}{\sin{\varphi}}.\label{lbe-rescaled}
\end{equation}
The general solution is
\begin{equation}
 \xi(y,\varphi) = a(\varphi)e^{-\frac{\gamma_i y}{\sin{\varphi}}} + \int_0^y dy^\prime e^{-\frac{\gamma_i(y-y^\prime)}{\sin{\varphi}}} \frac{\chi(y^\prime, \varphi)}{\sin{\varphi}}.\label{right-part}
\end{equation}

Plugging the expression for $\varepsilon(y,\varphi)$ into Eqs.~\eqref{left-boundary} and~\eqref{right-boundary} and solving for $a(\varphi)$, we find after some algebra~\cite{supplement}:
\begin{equation}
 \xi(y, \varphi) = \int_{-1/2}^{1/2} dy^\prime G_\varphi(y,y^\prime) \chi(y^\prime, \varphi),\label{right-integral}
\end{equation}
where
\begin{multline}
 G_\varphi(y,y^\prime) = \frac{e^{-\frac{\gamma_i(y-y^\prime)}{\sin{\varphi}}}}{\sin{\varphi}\left(1-r_\varphi e^{-\frac{\gamma_i}{\sin{\varphi}}}\right)}\left[\theta(y-y^\prime)  \right.\\+\left. r_\varphi\theta(y^\prime-y)  e^{-\frac{\gamma_i}{\sin{\varphi}}}\right],\qquad \varphi\in [0,\pi],\label{right-kernel}
\end{multline}
\begin{equation}
 G_\varphi(y,y^\prime) = G_{2\pi-\varphi}(-y,-y^\prime),\qquad \varphi\in[\pi,2\pi].\label{left-kernel}
\end{equation}

\textit{Perturbative analysis for the total current\/.}--- We now employ Eq.~\eqref{right-integral} to systematically compute the corrections to the solution of Eq.~\eqref{final-lbe} in powers of the electron-electron scattering operator. Once $\xi(y,\varphi)$ is known to the desired accuracy, we find the local current density as
\begin{multline}
 j(y) = 2\int \frac{d^2k}{(2\pi)^2} e\delta f v \cos(\varphi) = \rho_{2D} e^2 EW v_F\\ \times \frac{1}{2\pi}\int_0^{2\pi} d\varphi \xi(y, \varphi) \cos{\varphi}
\end{multline}
and the total current
\begin{equation}
 I = W\int_{-1/2}^{1/2} dy j(y) = \frac{E\rho_{2D}}{2}e^2W^2 v_F\times \Xi,
\end{equation}
where
\begin{equation}
 \Xi = \frac{1}{\pi} \int_{-1/2}^{1/2}dy \int_{0}^{2\pi} d\varphi\xi(y, \varphi) \cos{\varphi}.\label{Xidef}
\end{equation}
Here, $\rho_{2D}$ is the density of states at the Fermi energy. For the parabolic spectrum with effective mass $m^*$, $\rho_{2D} = m^*/(\pi\hbar^2)$.

To find the zeroth approximation for $\xi(y,\varphi)$, we neglect the electron-electron collision completely and substitute $\chi_0(y,\varphi) = \cos{\varphi}$ into Eq.~\eqref{right-integral}:
\begin{equation}
 \xi_0(y, \varphi) = \frac{\cos{\varphi}}{\gamma_i} \left[1 - h_\varphi e^{-\frac{\gamma_i}{2|\sin{\varphi}|}} e^{-\frac{\gamma_iy }{\sin{\varphi}}}\right],\label{xi-zero}
\end{equation}
where
\begin{equation}
 h_\varphi = \frac{1-r_\varphi}{1-r_\varphi e^{-\frac{\gamma_i}{|\sin{\varphi}|}}}\label{boundary-func}
\end{equation}
Here, we assume $r_{2\pi-\varphi} = r_\varphi$.
Plugging the result into Eq.~\eqref{Xidef}, we find after some algebra
\begin{equation}
 \Xi_0 = \frac{1}{\gamma_i} - \frac{4}{\pi\gamma_i^2} \int_{0}^{\pi/2}d\varphi h_\varphi\cos^2{\varphi} \sin{\varphi} \left(1-e^{-\frac{\gamma_i}{\sin{\varphi}}}\right).\label{Xi-zero}
\end{equation}

For the first-order contribution, we take the first-order correction to $\chi(y,\varphi)$ as
\begin{equation}
 \delta^{(1)}\chi(y, \varphi) = \int_{0}^{2\pi} d\varphi^\prime \tilde I_{ee}(\varphi-\varphi^\prime) \xi_0(y, \varphi^\prime),
\end{equation}
plug it into Eq.~\eqref{right-integral}, and subsequently into Eq.~\eqref{Xidef}.
After some tedious but straightforward integration~\cite{supplement}, we arrive at the main analytical result of the paper
\begin{multline}
\delta^{(1)}\Xi = \frac{4\gamma_{ee}^{(\mathrm{even})}}{\pi\gamma_i^2} \left[\frac{2}{\pi}\sum_{k=0}^{k^*}\frac{\Delta \gamma_k}{\gamma_{ee}^{(\mathrm{even})}} K_k(\gamma_i,\{r_\varphi\}) \right.\\\left. \vphantom{\sum_{k=0}^{k^*}}- J(\gamma_i,\{r_\varphi\})\right],\label{Xi-first-order}
\end{multline}
where
\begin{multline}
 K_k(\gamma_i,\{r_\varphi\}) = \frac{1}{4}\iint\limits_0^{\pi} d\varphi d\varphi^\prime h_\varphi \cos{\varphi} h_{\varphi^\prime}\cos{\varphi^\prime}
 \\ \times
 \cos\left[(2k+1)(\varphi - \varphi^\prime)\right]
 \left\{\frac{e^{-\frac{\gamma_i}{\sin{\varphi^\prime}}} - e^{-\frac{\gamma_i}{\sin{\varphi}}}}{\frac{\gamma_i}{\sin{\varphi}} - \frac{\gamma_i}{\sin{\varphi^\prime}}}\right.\\ \left.+
 \frac{1 - e^{-\frac{\gamma_i}{\sin{\varphi}} - \frac{\gamma_i}{\sin{\varphi^\prime}}}}{\frac{\gamma_i}{\sin{\varphi}}+\frac{\gamma_i}{\sin{\varphi^\prime}}}
 \right\}\label{Kdef}
\end{multline}

and
\begin{equation}
 J(\gamma_i,\{r_\varphi\}) = \int\limits_{0}^{\pi/2} d\varphi \cos^2{\varphi} h_\varphi^2 e^{-\frac{\gamma_i}{\sin{\varphi}}}.\label{Jdef}
\end{equation}
The functions $K_k(\gamma_i,\{r_\varphi\})$ and $J(\gamma_i,\{r_\varphi\})$ depend on the angular-dependent reflectivity parameter $r_\varphi$ through $h_\varphi$.
In principle, Eq.~\eqref{Xi-first-order} is quite general in the sense that we can plug any dependence of the odd scattering rates we want into it.

The sign of the correction to the conductance is directly influenced by the deviations $\Delta\gamma_k$.
At low temperatures, we can assume that $\gamma_{ee}^{2k+1}\ll \gamma_{ee}^{(\mathrm{even})}$ implying $\Delta \gamma_k/\gamma_{ee}^{(\mathrm{even})} = 1$.
In this case, we use
\begin{equation}
 \sum_{k=0}^{+\infty} \frac{\cos{\left[(2k+1)(\varphi-\varphi^\prime)\right]}}{\pi} = \frac{\delta_c(\varphi-\varphi^\prime) - \delta_c(\varphi-\varphi^\prime + \pi)}{2}
\end{equation}
to sum up the contributions of the odd harmonics and obtain
\begin{equation}
 \delta^{(1)}\Xi = \frac{2\gamma_{ee}^{(\mathrm{even})}}{\pi\gamma_i^2} \int_{0}^{\pi/2} d\varphi h_\varphi^2 \cos^2{\varphi} P(\gamma_i/\sin{\varphi}),\label{Xi-linear}
\end{equation}
\begin{equation}
 P(x) = e^{-x}\left[\cfrac{\sinh{x}}{x}-1\right]\geqslant 0.
\end{equation}
The equality is only valid for $x=0$. Since $\gamma_i/|\sin{|\varphi}|\geqslant \gamma_i$, it is never reached in the integrand of Eq.~\eqref{Xi-linear}. Therefore, the integral is strictly positive. Since $\gamma_{ee}^{(\mathrm{even})}\propto T^2$, the conductivity $G\propto \Xi$ initially grows with the square of the temperature.

At high temperature there are no deviations $\Delta\gamma_{k>0}\equiv 0$ and only the $k=0$ contribution survives in Eq.~\eqref{Xi-first-order}. For $\gamma_i\lesssim 1$, the term $J(\gamma_i, \{r_\varphi\})$ dominates and the correction $\delta^{(1)}\Xi$ is then negative. Note that if we additionally take the boundary conditions to be fully diffusive $r_\varphi\equiv0$, Eq.~\eqref{Xi-first-order} reproduces the result of Ref.~\cite{dejong_1995}.

As the temperature is increased from zero, both $k^*$ and $\Delta \gamma_k/\gamma_{ee}^{(\mathrm{even})}$ become smaller and Eq.~\eqref{Xi-first-order} interpolates between the two limits we have just considered, resulting in the appearance of a peak.
This argument is insensitive to the finer details of the odd scattering rate behaviour as long as the number of long-lived harmonics decays with temperature~\footnote{We have also checked it numerically in the broader range of temperatures. See the Supplementary Material~\cite{supplement} for more details.}.

\begin{figure}[t]
\begin{center}
\includegraphics[width=245pt]{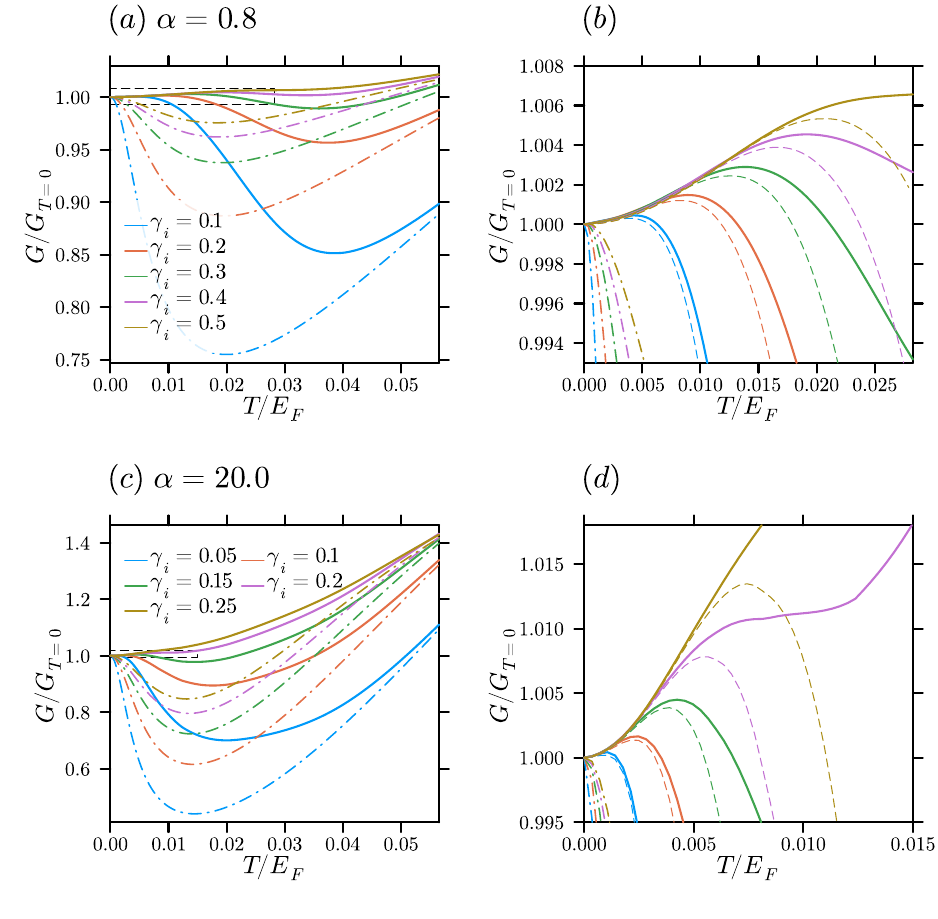}
\caption{Normalized conductance $G/G_{T=0}$ as a function of temperature
for different fixed values of the impurity scattering rate $\gamma_i$: $(a,b)$ for the Soffer parameter $\alpha=0.8$; $(c,d)$ for the Soffer parameter $\alpha=20.0$. Panels $(a,c)$ compare the computed conductances for the collision kernel with the scattering times given by Eqs.~\eqref{tau-even} and~\eqref{tau-odd} (solid lines) and for the collision kernel with the mode-independent scattering times $1/\tau_{ee}^{(n\geqslant2)} \equiv 1/\tau_{ee}^{(\mathrm{even})}$ (dash-dotted lines).
Panel $(b)$ zooms in on the part of panel $(a)$ highlighted by dashed rectangle and compares solid lines of panel (a) with the perturbative result of Eq.~\eqref{Xi-first-order} obtained using Eqs.~\eqref{tau-even} and~\eqref{tau-odd} (dashed lines). Analogously, panel $(d)$ zooms in on panel $(c)$. The corresponding values of $\gamma_i$ are colour-coded.
}
\label{fig-li}
\end{center}
\end{figure}

\begin{figure}[t]
\begin{center}
\includegraphics[width=245pt]{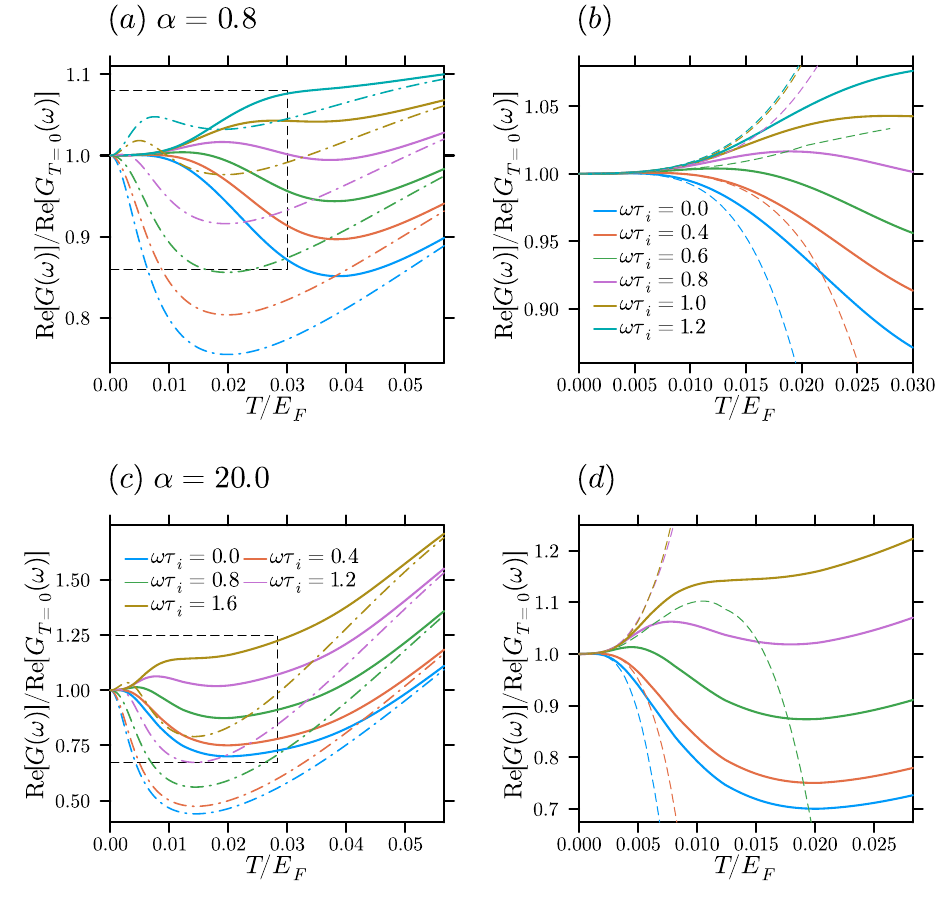}
\caption{Normalized real part of the conductance $\re{[G(\omega)]}/\re{[G_{T=0}(\omega)]}$ as a function of temperature for different fixed values of the ac drive frequency $\omega$: $(a,b)$ for the Soffer parameter $\alpha=0.8$; $(c,d)$ for the Soffer parameter $\alpha=20.0$. Panels $(a,c)$ compare the computed conductances for the collision kernel with the scattering times given by Eqs.~\eqref{tau-even} and~\eqref{tau-odd} (solid lines) and for the collision kernel with the mode-independent scattering times $1/\tau_{ee}^{(n\geqslant2)} \equiv 1/\tau_{ee}^{(\mathrm{even})}$ (dash-dotted lines).
Panel $(b)$ zooms in on the part of panel $(a)$ highlighted by dashed rectangle and compares solid lines of panel (a) with the perturbative result of Eq.~\eqref{Xi-first-order} obtained using Eqs.~\eqref{tau-even} and~\eqref{tau-odd} (dashed lines). Analogously, panel $(d)$ zooms in on panel $(c)$. The corresponding values of $\omega$ are colour-coded. The impurity scattering rate was fixed at $\gamma_i = 0.05$.}
\label{fig-omega}
\end{center}
\end{figure}

\textit{Numerical results and Discussion\/.}--- In order to study the conductance behaviour in the larger temperature range, we discretize Eq.~\eqref{final-lbe} together with the boundary conditions~\eqref{left-boundary} and~\eqref{right-boundary} and solve the resulting linear system of equations numerically. The results of the computations are presented in Fig.~\ref{fig-li}. To emphasize the features related to the long-lived odd harmonics, we compare the results obtained for the collision kernel with the relaxation times~\eqref{tau-even} and~\eqref{tau-odd} with the results obtained for the collision kernel with mode-independent relaxation times $1/\tau_{ee}^{(n\geqslant2)} \equiv 1/\tau_{ee}^{(\mathrm{even})}$. In addition to that we plot the results obtained using Eq.~\eqref{Xi-first-order}.

To discuss the numerics, it is convenient to rewrite Eqs.~\eqref{tau-even} and~\eqref{tau-odd} as
\begin{equation}
 \gamma_{ee}^{(\mathrm{odd})} = a^{-1} \left(\gamma_{ee}^{(\mathrm{even})}\right)^2. \label{func-dependence}
\end{equation}
where, up to logarithmic corrections~\cite{giuliani_1982, zheng_1996} (see Eq.~\eqref{tau-even}),
\begin{equation}
 a \cong \frac{\pi^2WE_F}{2v_F h} = \frac{\pi k_F W}{8}.\label{adef}
\end{equation}
We have chosen $a^{-1}=1/5000$ for our simulations. For the boundary conditions we have used the Soffer model~\cite{soffer_1967} $r_\varphi=e^{-\alpha\sin^2{\varphi}}$ with either $\alpha=0.8$ or $\alpha=20.0$.

The crucial effect of the long-lived odd harmonics is the appearance of the anomalous Knudsen peak preceding the usual Gurzhi dip and the shift of the Gurzhi dip to higher temperatures. Increasing the impurity scattering rate $\gamma_i$ moves the Gurzhi dip up, which makes this peak at smaller temperatures more pronounced.
More diffusive boundaries better mix the harmonics. As a result, the Gurzhi dip shifts to the left. Another interesting feature in the more diffusive case is the higher sensitivity of the results to the value of $\gamma_i$.
In stark contrast, the conductance curves obtained with mode-independent scattering times never exhibit a peak at small temperatures, even when the $\gamma_i$ is increased further and the Gurzhi dip vanishes being replaced by monotonous growth~\cite{supplement}.

In experiments, the temperature of the electrons can be controlled by applying a bias current~\cite{dejong_1995}.
The crucial parameter for the observation of the anomalous Knudsen peak is the dimensionless constant $a$.
Our numerical simulations suggest, that one needs $a\gg 100$ for the peak to be visible.
In the case of the more specular boundary conditions, we can understand it from the following qualitative argument.
We expect the position of the Gurzhi dip to be controlled by the slowest decay channel corresponding to the third harmonic. The minimum of the resistance occurs when the respective scattering length $l_{ee}^{(3)}$ is of the order of the strip width $W$:
\begin{equation}
 \gamma_{ee}^{(3)} = \frac{W}{l_{ee}^{(3)}}=
 \min\left\{81\left(\gamma_{ee}^{(\mathrm{even})}\right)^2/a, \gamma_{ee}^{(\mathrm{even})}\right\} \sim 1.
\end{equation}
For $a\lesssim81$, the long-lived harmonics disappear even before the minimum of the conductance is reached. It means that the Gurzhi dip stays unaffected and the peak disappears~\footnote{When the Gurzhi dip shifts to the left, it eats away the width of the peak.}.

For comparison, in the experiment of Ref.~\cite{ginzburg_2023}, the authors consider AlGaAs/GaAs heterostructures with low temperature scattering length $l_i = 60\ \mathrm{\mu m}$ and electron density $n=2.7\times 10^{11}\ \mathrm{cm}^{-2}$, corresponding to a Fermi wave-vector of $k_F = \sqrt{2\pi n} \approx 1.3\times 10^6\ \mathrm{cm}^{-1}$. For a strip of the width $W=15\ \mathrm{\mu m}$, that would give $\gamma_i = W/l_i=0.25$ and $a\sim800$.
These numbers suggest that the anomalous Knudsen peak can be observed in state-of-the-art experiments on clean 2D electron systems.

In light of the sensitivity of the conductance curves to the value of $\gamma_i$, it is advantageous to have some control over this parameter in the experiment, which does not require engineering different samples. To some extent, this can be achieved if we use ac external fields instead of dc fields to induce the electron flow in the strip. On the level of the linearized Boltzmann equation~\eqref{lbe-raw}, having an ac drive amounts to the appearance of an additional term $-i\omega f$ on the right-hand side, which can be absorbed into the renormalization of the dimensionless scattering rate $\gamma_i\rightarrow \gamma_i - i\omega W/v_F$.
We have studied the influence of the frequency of the ac drive on the conductance curves numerically. The results are displayed in Fig.~\ref{fig-omega}. Similar to changing $\gamma_i$ directly, increasing the frequency pushes the Gurzhi dip upwards, making the anomalous Knudsen peak more pronounced. In contrast, for mode-independent relaxation times, we notice that the corresponding conductance curves also develop a peak at smaller temperatures. This however becomes noticeable only at higher freguencies $\omega\tau_i\gtrsim 0.8$. Thus, at smaller frequencies the anomalous Knudsen peak remains the unique signature of the long-lived odd harmonics.

\textit{Summary\/.}--- In this paper, we have studied the temperature dependence of the conductance in a long narrow strip and have demonstrated that taking into account the long-lived chararacter of the odd harmonics of the electron distribution function leads to the appearance of the anomalous Knudsen peak preceding the usual Gurzhi dip. The simultaneous observation of both these features in the conductance curves
constitutes a particular signature of the long-lived modes in 2D electron transport at low temperatures.

\textit{Acknowledgements\/.}--- We would like to thank Prof. Dr. Laurens W. Molenkamp and Dr. Pavlo Pyshkin and for the fruitful discussions. This work was supported by the DFG-SFB 1170 (Project-ID: 258499086) and EXC2147 ctd.qmat (Project-ID 390858490).

\nocite{fuchs_1938}

\bibliography{AnomalousKnudsen.bib}
\onecolumngrid
\clearpage

\begin{center}
  \bf\large
  Supplementary Material\\ for the article\\ ``Anomalous Knudsen effect signaling long-lived modes in 2D electron gases''
\end{center}\bigskip

\twocolumngrid

\setcounter{figure}{0}
\setcounter{equation}{0}
\setcounter{table}{0}
\setcounter{section}{0}

\renewcommand{\thefigure}{S\arabic{figure}}

\renewcommand{\theequation}{S\arabic{equation}}

\renewcommand{\thetable}{S\arabic{table}}

\renewcommand{\thesection}{S\Roman{section}}

All equation numbers, figure numbers and reference numbers without prefix ``S'' refer to the respective numbers in the main article.

\section{Derivation of the integral equation.\label{int-deriv}}

In order to condense the notation, let us define
\begin{equation}
 g(x, \varphi) = \int_0^y dy^\prime e^{-\frac{\gamma_i(y-y^\prime)}{\sin{\varphi}}}\frac{\chi(y^\prime,\varphi)}{\sin{\varphi}}.
\end{equation}
This is simply the integral term in the Equation~\eqref{right-part}. In the following we will treat the right- and left-moving electrons separately: we limit $\varphi\in[0,\pi]$ and consider $a(\varphi)$ and $a(2\pi-\varphi)$ as two independent functions.

Since $\left(-\xi(y,\pi-\varphi)\right)$ also satisfies Eq.~\eqref{final-lbe}, the solutions have the symmetry $\xi(y,\varphi) = -\xi(y,\pi-\varphi)$, due to which Eqs.~\eqref{left-boundary} and~\eqref{right-boundary} simplify to the form of the Fuchs boundary condition~\cite{fuchs_1938}:
\begin{eqnarray}
 \xi\left(-\frac{1}{2}, \varphi\right) & = & r_\varphi \xi\left(-\frac{1}{2},2\pi-\varphi\right),\label{fuchs-left}\\
 \xi\left(\frac{1}{2}, 2\pi-\varphi\right) & = & r_\varphi \xi\left(\frac{1}{2},\varphi\right).\label{fuchs-right}
\end{eqnarray}

Substituting everything into the simplified boundary conditions~\eqref{fuchs-left} and~\eqref{fuchs-right} we obtain
\begin{multline}
 a(\varphi) e^{\frac{\gamma_i}{2\sin\varphi}} + g(-1/2,\varphi) \\ = r_\varphi a(2\pi-\varphi) e^{-\frac{\gamma_i}{2\sin{\varphi}}} + r_\varphi g(-1/2,2\pi-\varphi),
\end{multline}
\begin{multline}
 a(2\pi-\varphi) e^{\frac{\gamma_i}{2\sin{\varphi}}} + g(1/2, 2\pi-\varphi) \\= r_\varphi a(\varphi)e^{-\frac{\gamma_i}{2\sin{\varphi}}} + r_\varphi g(1/2, \varphi).
\end{multline}
Solving for $a(\varphi)$ we get
\begin{multline}
 \left [1 - r_\varphi^2 e^{-\frac{2\gamma_i}{\sin{\varphi}}}\right]a(\varphi) = -e^{-\frac{\gamma_i}{2\sin{\varphi}}}g(-1/2, \varphi) \\- r_\varphi e^{-\frac{3\gamma_i}{2\sin{\varphi}}} g(1/2, 2\pi-\varphi) + r_\varphi e^{-\frac{\gamma_i}{2\sin{\varphi}}} g(-1/2, 2\pi-\varphi) \\+ r_\varphi^2 e^{-\frac{3\gamma_i}{2\sin{\varphi}}} g(1/2, \varphi).
\end{multline}
Finally, substituting everything back into Eq.~\eqref{right-part}, we arrive at
\begin{multline}
 \xi(y, \varphi) = \frac{1}{1-r_\varphi^2e^{-\frac{2\gamma_i}{\sin{\varphi}}}}\left[\int_{-1/2}^y dy^\prime e^{-\frac{\gamma_i(y-y^\prime)}{\sin{\varphi}}}\frac{\chi(y^\prime, \varphi)}{\sin{\varphi}}\right.\\
 + r_\varphi^2 e^{-\frac{2\gamma_i}{\sin{\varphi}}}\int_{y}^{1/2}dy^\prime e^{-\frac{\gamma_i(y-y^\prime)}{\sin{\varphi}}}\frac{\chi(y^\prime,\varphi)}{\sin{\varphi}} \\
 \left.+ r_\varphi e^{-\frac{\gamma_i}{\sin{\varphi}}} \int_{-1/2}^{1/2}dy^\prime e^{-\frac{\gamma_i(y+y^\prime)}{\sin\varphi}}\frac{\chi(y^\prime, 2\pi-\varphi)}{\sin{\varphi}}\right].
\end{multline}
Every term in this expression has a clear physical meaning.
The right-hand side of Eq.~\eqref{lbe-rescaled} creates the deformation of the distribution function that is picked up and propagated by the electrons. A scattering on the impurity or a diffusive scattering at the boundary completely randomizes the direction of electrons velocity, hence
we only need to sum over the contributions of the nonscattered electron trajectories, which look like straight lines reflected at the boundaries.
For every such trajectory, the combination of $r_\varphi$-s and exponential factors takes into account the probability that the electron actually flew the trajectory without scattering.

The first term describes then the contribution of the right-moving electrons that arrive from the left. The right-moving electrons with $y^\prime>y$ reflect two times before arriving at the point with the coordinate $y$ and provide the second term. The third term is the contribution of the left-moving electrons, that need to reflect once to change the direction. Finally, the prefactor
\begin{equation}
 \frac{1}{1-r_\varphi^2e^{-\frac{2\gamma_i}{\sin{\varphi}}}} = 1 + r_\varphi^2e^{-\frac{2\gamma_i}{\sin{\varphi}}} + \left(r_\varphi^2e^{-\frac{2\gamma_i}{\sin{\varphi}}}\right)^2 + \dotsc
\end{equation}
takes into account the trajectories where the electron additionally travels the distance between the two edges back and forth multiple times while reflecting at the boundaries.

The expression can be further simplified if we notice that the problem has the mirror symmetry with respect to the central axis $y=0$ of the channel and therefore $\xi(-y,2\pi-\varphi) = \xi(y, \varphi)$ which is also satisfied by $\chi(y, \varphi)$. Transforming the third term with the help of this identity and factoring out the common denominator $\left(1 + r_\varphi e^{-\frac{\gamma_i}{\sin{\varphi}}}\right)$ we arrive at Eqs.~\eqref{right-integral} and~\eqref{right-kernel}. The kernel~\eqref{left-kernel} for the left-moving electrons follows then again from the same symmetry considerations. Indeed, if $\varphi\in[\pi,2\pi]$, then
\begin{multline}
 \xi(y, \varphi) = \xi(-y, 2\pi - \varphi) \\= \int_{-1/2}^{1/2} dy^\prime G_{2\pi-\varphi}(-y, y^\prime) \chi(y^\prime, 2\pi-\varphi)
 \\ =
 \int_{-1/2}^{1/2} dy^\prime G_{2\pi-\varphi}(-y, -y^\prime) \chi(y^\prime, \varphi).
\end{multline}

\section{Derivation of the first-order correction to the current.}

First of all, we use the mirror symmetry to simplify Eq.~\eqref{Xidef}:
\begin{multline}
 \Xi = \frac{1}{\pi}\int_{-1/2}^{1/2}dy \int_{0}^{\pi} d\varphi \cos{\varphi}\left[\xi(y, \varphi) + \xi(y, 2\pi-\varphi)\right]
 \\ =
 \frac{1}{\pi}\int_{-1/2}^{1/2}dy \int_{0}^{\pi} d\varphi \cos{\varphi}\left[\xi(y, \varphi) + \xi(-y, 2\pi-\varphi)\right]
 \\ =
 \frac{2}{\pi}\int_{-1/2}^{1/2} dy \int_{0}^{\pi} d\varphi \cos{\varphi}\xi(y, \varphi).
\end{multline}
Substituting here Eq.~\eqref{right-integral}
\begin{equation}
 \Xi = \frac{2}{\pi}\iint\limits_{-1/2}^{1/2} dy dy^\prime \int_{0}^{\pi}d\varphi \cos{\varphi} G_\varphi(y,y^\prime) \chi(y^\prime, \varphi),
\end{equation}
we can evaluate the outer integral in $y$ right away to obtain
\begin{equation}
 \Xi = \frac{2}{\pi\gamma_i} \int_{-1/2}^{1/2} dy^\prime \int_0^\pi d\varphi \cos{\varphi} G_\varphi(y^\prime) \chi(y^\prime, \varphi),
\end{equation}
\begin{equation}
 G_{\varphi}(y^\prime) = \gamma_i\int_{-1/2}^{1/2} dy G_\varphi(y,y^\prime) = 1 - h_\varphi e^{\frac{\gamma_i(y^\prime-1/2)}{\sin{\varphi}}}.
\end{equation}

Secondly, we remember that $\chi(y^\prime, \varphi)$ is the right-hand side of Eq.~\eqref{lbe-rescaled}:
\begin{multline}
 \Xi = \frac{2}{\pi\gamma_i} \int_{-1/2}^{1/2} dy^\prime \int_{0}^\pi d\varphi \cos^2{\varphi} G_\varphi(y^\prime)
 + \frac{2}{\pi\gamma_i}\\\times \int_{-1/2}^{1/2} dy^\prime \int_0^\pi d\varphi \int_{0}^{2\pi}d\varphi^\prime \cos{\varphi} G_\varphi(y^\prime) \tilde I_{ee}(\varphi-\varphi^\prime) \xi(y^\prime,\varphi^\prime)
 \\ = \Xi_0 + \delta \Xi.\label{Xi-transformed-raw}
\end{multline}
The first term gives exactly $\Xi_0$ from Eq.~\eqref{Xi-zero}.

The second term provides the correction to the current due to electron-electron collisions. Yet again we explicitly split the contributions from the right- and left-moving electrons in the integral over $\varphi^\prime$:
\begin{multline}
 \delta\Xi = \frac{2}{\pi\gamma_i} \int_{-1/2}^{1/2} dy^\prime \iint\limits_{0}^\pi d\varphi d\varphi^\prime \cos{\varphi} G_\varphi(y^\prime) \\\times\left[\tilde I_{ee}(\varphi-\varphi^\prime) \xi(y^\prime, \varphi^\prime) + \tilde I_{ee}(\varphi+\varphi^\prime) \xi(y^\prime, 2\pi-\varphi^\prime)\right]
 \\ = \frac{2}{\pi\gamma_i} \int_{-1/2}^{1/2} dy^\prime \iint\limits_{0}^\pi d\varphi d\varphi^\prime \cos{\varphi} \xi(y^\prime,\varphi^\prime)\\
 \times
 \left[G_\varphi(y^\prime) \tilde I_{ee}(\varphi-\varphi^\prime) + G_\varphi(-y^\prime) \tilde I_{ee}(\varphi+\varphi^\prime)\right].\label{Xi-transformed}
\end{multline}
In the first equality we have used the $2\pi$-periodicity of $\tilde I_{ee}(\varphi)$ to transform
\begin{equation}
 \tilde I_{ee}(\varphi - (2\pi-\varphi^\prime)) = \tilde I_{ee}(\varphi+\varphi^\prime).
\end{equation}
In the second equality we have changed the integration variable $y^\prime\rightarrow-y^\prime$ and used the mirror symmetry. If we want to find the first-order correction, we just need to subtstitute the zeroth-order approximation~\eqref{xi-zero} for $\xi(y^\prime, \varphi^\prime)$.

Now, there are several simplifying considerations that prove to be helpful.
The first one is related to the fourier series representation of the kernel~\eqref{deviation-form}. In the expansions of $\tilde I_{ee}(\varphi-\varphi^\prime)$ or $\tilde I_{ee}(\varphi+\varphi^\prime)$, we have the terms
\begin{multline}
 \frac{\Delta\gamma_k}{\pi}\cos[(2k+1)(\varphi\mp\varphi^\prime)] \\= \frac{\Delta\gamma_k}{\pi}\{\cos{[(2k+1)\varphi]} \cos{[(2k+1)\varphi^\prime]} \\\pm \sin{[(2k+1)\varphi]}\sin{[(2k+1)\varphi^\prime]}\}.
\end{multline}
The functions $\sin{[(2k+1)\varphi]}$ and $G_{\varphi}(y)$ are even on $[0,\pi]$ as the functions of $\varphi$, while $\cos{\varphi}$ and $\xi(y, \varphi)$ (and $\xi_0(y,\varphi)$ too) are all odd. Because of that the integrals with the sine-terms all fall out, and we can replace $-\tilde I_{ee}(\varphi-\varphi^\prime)$ with
\begin{multline}
 -\tilde I_{ee}(\varphi,\varphi^\prime) = \gamma_{ee}^{(\mathrm{even})}\left[\delta_c(\varphi-\varphi^\prime)-1/(2\pi)\right] \\- \sum_{k=0}^{k^*} \frac{\Delta\gamma_k}{\pi} \cos{\left[(2k+1)(\varphi)\right]\cos{[(2k+1)\varphi^\prime]}}\label{kernel-replaced}
\end{multline}
everywhere without changing the end result.

The second consideration is the fact that the electron collision term conserves momentum ($\gamma_{ee}^{(1)}=0$) and the number of particles ($\gamma_{ee}^{(0)}=0$) and therefore
\begin{equation}
 \int_{0}^{2\pi} d\varphi^\prime \tilde I_{ee}(\varphi-\varphi^\prime) \cos{\varphi^\prime} = \int_{0}^{2\pi} d\varphi^\prime \tilde I_{ee}(\varphi-\varphi^\prime) = 0,
\end{equation}
Because of that the first term in Eq.~\eqref{xi-zero} falls out when we substitute it into Eqs.~\eqref{Xi-transformed-raw} and~\eqref{Xi-transformed}:
\begin{multline}
 \delta^{(1)}\Xi = -\frac{2}{\pi\gamma_i^2} \int\limits_{-1/2}^{1/2} dy^\prime \iint\limits_0^\pi d\varphi d\varphi^\prime \cos{\varphi}\cos{\varphi^\prime} h_{\varphi^\prime} e^{-\frac{\gamma_i(1/2+y^\prime)}{\sin{\varphi^\prime}}}
 \\ \times \left[G_\varphi(y^\prime) \tilde I_{ee}(\varphi,\varphi^\prime) + G_\varphi(-y^\prime) \tilde I_{ee}(\varphi,-\varphi^\prime)\right].
\end{multline}
We can also ignore the constant $-\gamma_{ee}^{(\mathrm{even})}/(2\pi)$ in Eq.~\eqref{kernel-replaced}.

The last integration over $y^\prime$ is straightforward and one arrives in the end at Eq.~\eqref{Xi-first-order}. The only detail worth mentioning is that one actually gets factor $\cos{\left[(2k+1)(\varphi)\right]\cos{[(2k+1)\varphi^\prime]}}$ in place of $\cos[(2k+1)(\varphi-\varphi^\prime)]$ in Eq.~\eqref{Kdef}. However, we can add the product of sines back without changing the integrals to get Eq.~\eqref{Kdef}.

\section{Influence of the precise form of the odd scattering rates on the results.}

\begin{figure*}
 \includegraphics[width=450pt]{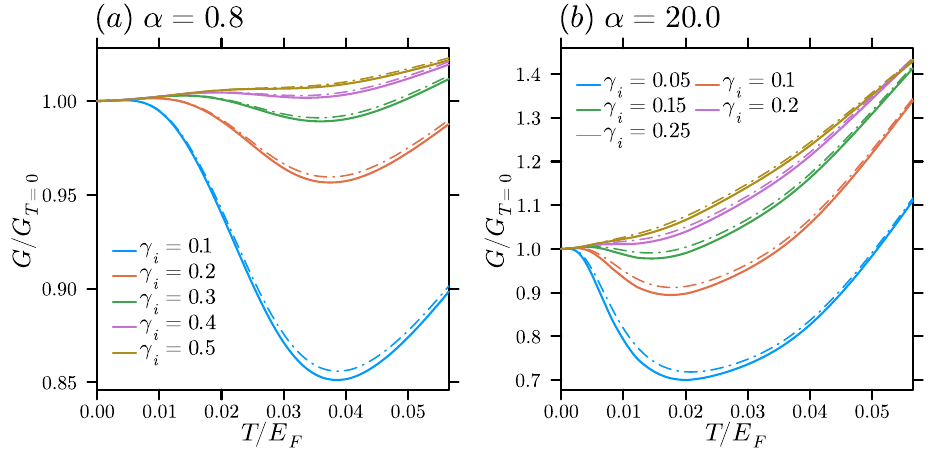}
 \caption{Normalized conductance $G/G_{T=0}$ as a function of temperature
for different fixed values of the impurity scattering rate $\gamma_i$ for the collision integral with the odd scattering rates given by Eq.~\eqref{tau-odd} (solid lines) and for smoothly saturating odd scattering rates $\gamma_\mathrm{ee}^{(2k+1)} = \gamma_{ee}^{(\mathrm{odd})}(2k+1)^4/\left[1 + (2k+1)^4\gamma_{ee}^{(\mathrm{odd})}/\gamma_{ee}^{(\mathrm{even})}\right]$ (dash-dotted lines): $(a)$ for Soffer parameter $\alpha=0.8$; $(b)$ for Soffer parameter $\alpha=20.0$. The difference between the solid line and the dashed-dotted line for a given choice of $\gamma_i$ is small.}
\label{smooth-saturation}
\end{figure*}

In the manuscript, we choose the form~\eqref{tau-odd} of the odd scattering rates, where the $(2k+1)^4$ growth with the order of the harmonic abruptly changes into constant behaviour, as soon as $\gamma_{\mathrm{ee}}^{(2k+1)}$ reaches the level of $\gamma_{\mathrm{ee}}^{(\mathrm{even})}$. We made this choice for simlpicity. In reality, the odd scattering rates should smoothly interpolate between the $T^4$ behaviour at small $k$ and $T^2$ behaviour at large $k$~\cite{hofmann_2023}.

However, the fine details of the odd scattering rate scaling are not crucial for the main effect discussed in the manuscript. It is only crucial that at small temperatures almost all odd harmonics are long-lived ($\Delta \gamma_k \approx \gamma_{\mathrm{ee}}^{(\mathrm{even})}$ for $k\geqslant 1$), while with growing temperature the number of long-lived harmonics quickly shrinks ($\Delta \gamma_k$ approach zero with growing $T$, which happens faster for larger $k$). Qualitatively, wee can justify it on the basis of the theoretical result~\eqref{Xi-first-order}, where we can substitute any behaviour of the odd scattering rates.

We have also tested the influence of the precise form of the odd scattering rates numerically. In Ref.~\cite{kryhin_2025}, an alternative form of the odd scattering rates has been suggested:
\begin{equation}
    \gamma_\mathrm{ee}^{(2k+1)} = \frac{\gamma_\mathrm{ee}^{(\mathrm{odd})}(2k+1)^4}{1 + \frac{\gamma_\mathrm{ee}^{(\mathrm{odd})}(2k+1)^4}{\gamma_\mathrm{ee}^{(\mathrm{even})}}}.\label{tau-odd-smooth}
\end{equation}
At small $k$, it agrees with Eq.~\eqref{tau-odd} and hence $\gamma_\mathrm{ee}^{(2k+1)}\propto T^4$. At large $k$, $\gamma_\mathrm{ee}^{(2k+1)}=\gamma_\mathrm{ee}^{(\mathrm{even})}\propto T^2$. Overall, it has the desired feature, that it smoothly interpolates between the two scaling regimes at small and large $k$.
We compare the conductance curves $G(T)$ obtained numerically with the odd scattering rates described by Eq.~\eqref{tau-odd} and Eq.~\eqref{tau-odd-smooth} in Fig.~\ref{smooth-saturation}. As we see, the differences are rather small, which supports our argument.\bigskip

\section{Behaviour of the conductance curves at large impurity scattering rates $\gamma_i$ in the case of the collision kernel with mode-independent scattering rates.}

\begin{figure*}
 \includegraphics[width=450pt]{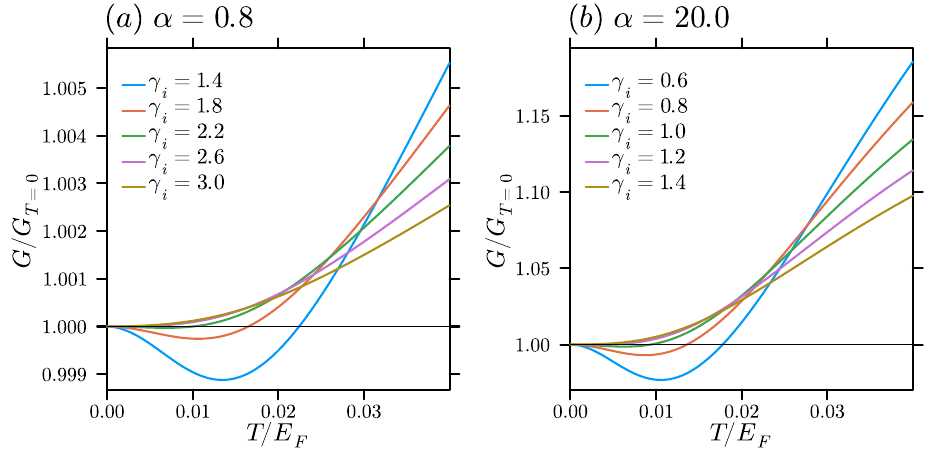}
 \caption{Normalized conductance $G/G_{T=0}$ as a function of temperature
for different fixed values of the impurity scattering rate $\gamma_i$ for the collision integral with collision kernel with the mode-independent scattering times $1/\tau_{ee}^{(n\geqslant2)} \equiv 1/\tau_{ee}^{(\mathrm{even})}$}: $(a)$ for Soffer parameter $\alpha=0.8$; $(b)$ for Soffer parameter $\alpha=20.0$. In both cases, we see that for large enough $\gamma_i$, the Gurzhi dip vanishes and the conductance simply grows monotonously with the temperature. Note that in the vicinity of such critical $\gamma_i$ when the behaviour switches, the conductance obtained with the mode-independent scattering rates never exhibits a peak at small temperatures.
\label{critical-gammai}
\end{figure*}

Analyzing Fig.~\ref{fig-li}, we make the following observation about the behaviour of the conductance curves obtained with the collision kernel taking into account the long-lived harmonics as $\gamma_i$ is increased: The Gurzhi dip is pushed up, while the anomalous Knudsen peak gets more pronounced. For larger values of $\gamma_i$, the Gurzhi dip vanishes and the conductance exhibits monotonous growth with temperature.

The conductance curves obtained with the collision kernel with mode-independent scattering rates demonstrate similar behaviour: The Gurzhi dip is pushed up. In principle, we could worry that as $\gamma_i$ is increased, the conductance curves obtained with mode-independent scattering rates also develop a peak at small temperature. If that were the case, it would completely invalidate the argument that the anomalous Knudsen peak is a clear sign of long-lived harmonics. However, as our numerical results presented in Fig.~\ref{critical-gammai} demonstrate, the conductance curves obtained with mode-independent scattering rates never develop such peak. As $\gamma_i$ is increased, decrease$\rightarrow$growth behaviour is simply replaced by monotonous growth with temperature.

The critical value of $\gamma_i$ for which the switch in behaviour occurs can be determined from Eq.~\eqref{Xi-first-order}. We just neeed to plug in the mode-independent odd scattering rates and look for the value of $\gamma_i$, for which the correction changes sign.

\end{document}